\documentclass[conference]{IEEEtran}
\IEEEoverridecommandlockouts
\usepackage{cite}
\usepackage{amsmath,amssymb,amsfonts}
\usepackage{algorithmic}
\usepackage{graphicx}
\usepackage{textcomp}
\usepackage{subfigure}
\usepackage{float}
\usepackage{xcolor}
\usepackage{enumerate}
\usepackage{graphicx}
\usepackage{subcaption}
\newtheorem{theorem}{Theorem}

\newtheorem{lemma}{Lemma}

\usepackage{geometry}
\allowdisplaybreaks[4]
\makeatletter
\renewcommand{\maketag@@@}[1]{\hbox{\m@th\normalsize\normalfont#1}}%
\makeatother
\begin{document}
	\title{Coding for Quasi-Static Fading Channel with Imperfect CSI at the Transmitter and Quantized Feedback}

\author{\IEEEauthorblockN{
	    Yuhan~Yang\IEEEauthorrefmark{1},
		Mei~Han\IEEEauthorrefmark{1},
		Haonan~Zhang\IEEEauthorrefmark{1},
    	Haoheng~Yuan\IEEEauthorrefmark{6},
		Fan~Cheng\IEEEauthorrefmark{4},
		Bin~Dai\IEEEauthorrefmark{1}\IEEEauthorrefmark{2}\IEEEauthorrefmark{3}}
	\IEEEauthorblockA{
		\IEEEauthorrefmark{1}
		School of Information Science and Technology, Southwest Jiaotong University, Chengdu, 610031, China.}
	\IEEEauthorblockA{
		\IEEEauthorrefmark{2}
		Chongqing Key Laboratory of Mobile Communications Technology, Chongqing, 400065, China.}
	\IEEEauthorblockA{
		\IEEEauthorrefmark{3}
		Provincial
		Key Lab of Information Coding and Transmission, Southwest Jiaotong
		University, Chengdu, 611756, China.}
	\IEEEauthorblockA{
		\IEEEauthorrefmark{4}
		Department of Computer Science and Engineering, Shanghai Jiao Tong University, Shanghai, 200241, China.}
	\IEEEauthorblockA{
	\IEEEauthorrefmark{6}
		School of Information Engineering and Automation, Kunming University of Science and Technology, Kunming, 650500, China.\\		
		yangyvhan@my.swjtu.edu.cn, 13880326688@163.com, zhanghaonan@my.swjtu.edu.cn, \\
		20240066@kust.edu.cn, chengfan@cs.sjtu.edu.cn,  daibin@home.swjtu.edu.cn.}
}

	\maketitle
	
	\begin{abstract}
The classical Schalkwijk-Kailath (SK) scheme for the additive Gaussian noise channel with noiseless feedback is highly efficient since its coding complexity is extremely low and the decoding error doubly exponentially decays as the coding blocklength tends to infinity. However,
its application to the fading channel with imperfect CSI at the transmitter (I-CSIT) is challenging since the SK scheme is sensitive to the CSI. In this paper, we investigate how to design SK-type scheme
for the quasi-static fading channel with I-CSIT and quantized feedback. By introducing modulo lattice function and an auxiliary signal into the SK-type encoder-decoder of the transceiver,
we show that the decoding error caused by the I-CSIT can be perfectly eliminated, resulting in the success of designing SK-type scheme for such a case.
The study of this paper provides a way to design efficient coding scheme for fading channels in the presence of imperfect CSI and quantized feedback.
	\end{abstract}
	
	\begin{IEEEkeywords}
	Fading channel, imperfect CSI, quantized feedback,  Schalkwijk-Kailath scheme.
	\end{IEEEkeywords}
	\section{Introduction}\label{sec1}
	
	Ultra-reliable low-latency communication (URLLC) \cite{fbl2} is the key to modern wireless communication since it supports many
	critical services requiring high level of reliability and
	low latency, such as unmanned
	aerial vehicle (UAV) communication network \cite{UAV}, Vehicle-to-Everything (V2X) communication \cite{V2X}, etc. Finite blocklength (FBL) coding \cite{fbl1} provides a promising approach to achieve URLLC, and among which is the Schalkwijk-Kailath (SK) scheme \cite{JS7} for the additive white Gaussian noise (AWGN) channel with perfect feedback. The SK scheme is an iterative coding scheme which depends on the receiver's minimum mean square estimation (MMSE) about its previous time's estimation error known and sent by the transmitter. Comparing with the well-known linear block codes, the decoding error of the SK scheme decreases as a second-order exponential in the coding blocklength, which indicates that for fixed decoding error, the coding blocklength of the SK scheme is much shorter.

\cite{fbl1} showed that the SK scheme almost approaches the FBL capacity of the AWGN channel with feedback, and in recent years, the SK scheme has been further developed in many cases.
Specifically, \cite{JS10} extended the classical SK scheme to the AWGN channel with AWGN feedback by introducing a modulo-lattice operation to mitigate the impact of feedback channel noise on the performance of the SK scheme. In parallel, \cite{JS11} studied the AWGN channel with quantized feedback, where the receiver's signal is quantized by its own quantizer before being fed back to the transmitter, and a feedback control based SK-type scheme was proposed for such a model. Besides this, \cite{Markov} and \cite{MIMO} respectively extended the SK scheme to the fading and MIMO cases, and both of them assume that perfect channel state information (CSI) is available at the transceiver.

However, in practical wireless systems, the receiver is often assumed to be able to obtain the perfect CSI as long as the training sequence is sufficiently long \cite{H_est}, and through
a quantized feedback channel (QFC), the transmitter only gets imperfect CSI caused by the quantized noise.
The imperfect CSI at the transmitter (I-CSIT) and the QFC lead to the fact that it is difficult to design an SK-type scheme for such a case since the SK scheme is
sensitive to the CSI and channel output feedback.
	
In this paper, we aim to extend the classical SK scheme to the quasi-static fading channel with I-CSIT and QFC. By introducing modulo lattice function and
	an auxiliary signal into the SK-type encoder-decoder of the transceiver, we show that the decoding error caused by I-CSIT can be perfectly eliminated, resulting in the success of
designing SK-type scheme for such a case.
	
	\begin{figure}[htb]
		\centering
		\includegraphics[scale=0.2]{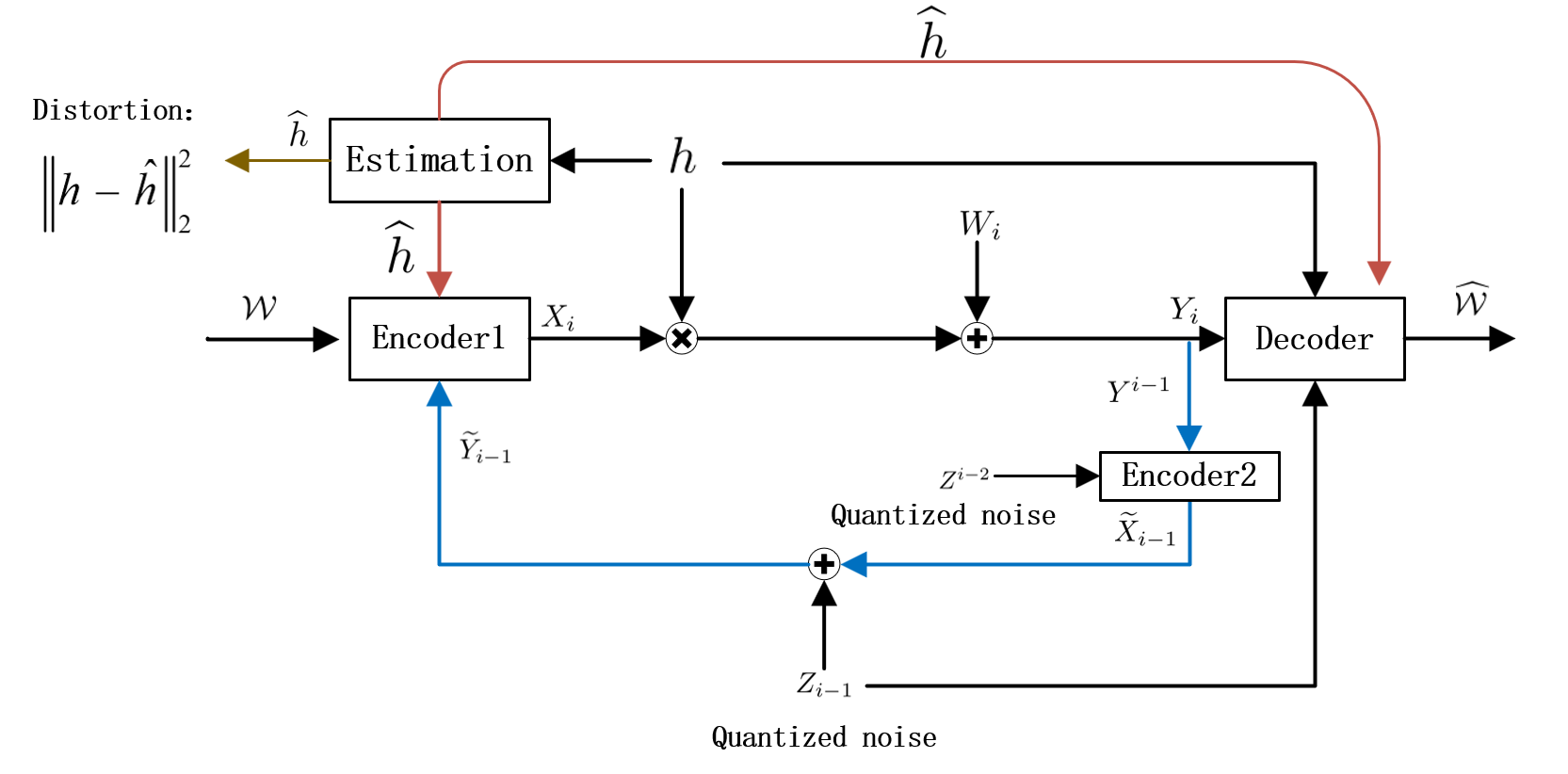}
		\caption{Quasi-static fading channel with I-CSIT and quantized feedback}
		\label{fig1}
	\end{figure}
	\section{Fading Channel with I-CSIT and QFC}\label{s1}
	\subsection{Model Formulation and Main Results}

The quasi-static fading channel with I-CSIT and QFC is shown in Figure \ref{fig1}, and at
time $i\in\left\lbrace 1,2,\dots,N\right\rbrace $, the channel inputs and outputs are given by

	\begin{equation}\label{model}
		\begin{array}{l}
			Y_i=hX_i+W_i, \\
\widetilde{Y}_{i}=\widetilde{X}_{i}+Z_{i},	
		\end{array}
	\end{equation}
	where $X_i$ is the channel input at time $i$ subject to an average power constraint {\small $\frac{1}{N}\sum\nolimits_{i=1}^{N}E({X}_{i}^2)\leq P$}, $h$ is a fixed non-zero channel fading coefficient
	\footnote{Such an assumption holds for the scenario with large coherence bandwidth or coherence time, e.g., indoor scenarios with small delay and Doppler spread \cite{H_1}~-\cite{H_2}.},
	which is perfectly known by the receiver,
	$W_i\sim\mathcal{N}(0,\sigma^{2})$ is the Gaussian noise which is
	independent and identically distributed (i.i.d.) across the time index $i$,
	$\widetilde{X}_i$ is the receiver's feedback codeword with average power constraint
	{\small $\frac{1}{N}\sum\nolimits_{i=1}^{N}E(\widetilde{X}_{i}^2)\leq \widetilde{P}$}, $Y_i$
	and $\widetilde{Y}_i$ are the outputs of the forward and feedback channels, respectively, and
	$Z_i=\mathbb{M}_{2\sigma_z}[\widetilde{X}_{i}]$ is the QFC noise which is generated by the local quantizer at the receiver \cite{JS11}, $\mathbb{M}_{2\sigma_z}[\cdot]$ is the lattice $\wedge_{2\sigma_z}$ in $\mathbb{R}$ spanned by a constant  $G=2\sigma_z$ such that
	\begin{equation}
		\wedge_{2\sigma_z}=\left\lbrace t=Ga:a\in \mathbb{Z}\right\rbrace.
	\end{equation}
	Denote the nearest neighbor quantizer associated with the lattice $\wedge_{2\sigma_z}$
	by
	\begin{equation}
		\mathbf{Q}_{\wedge_{2\sigma_z}}(x)\stackrel{\text {def}}{=}\text{arg} \min\limits_{t\in \wedge_{2\sigma_z}}\|x-t\|,
	\end{equation}
	where $\mathbb{M}_{\wedge_{2\sigma_z}}[\cdot]$ is the modulo-$\wedge_{2\sigma_z}$ function \cite{JS10} which is defined as
	\begin{equation}\label{sigma_z}
		\mathbb{M}_{\wedge_{2\sigma_z}}[x] \stackrel{\text {def}}{=} x-\mathbf{Q}_{\wedge_{2\sigma_z}}(x).
	\end{equation}
	The size of $\sigma_z$ measures the fineness of the quantizer, and  (\ref{sigma_z}) indicates that $\Pr\left\lbrace |Z_i|\leq\sigma_z\right\rbrace =1$.
	
	The signal-to-noise ratios (SNR) of the forward channel is denoted as	{\small $\text{SNR}\mathop{=}\limits^{\rm{def}}\frac{P}{\sigma^2}$}.

		%

	Define $\widehat{h}$ as the transmitter's estimation about $h$ via channel feedback, and
	the corresponding distortion of the estimation is denoted by the norm-bounded mode \footnote{In channel with quantized feedback \cite{JS11}, the norm-bounded model is commonly used as a measure of estimation error.
}
	\begin{equation}
		\bigtriangleup\mathop{=}\limits^{\rm{def}} \|h-\hat{h}\|_2. 	
	\end{equation} 			
	\emph{Channel coding}:
	\begin{itemize}
		\item 	The message $W$ is uniformly drawn from $\mathcal{W} = \left\lbrace 1, 2, . . . , M\right\rbrace $.
		\item 	At time index $i$ ($i \in\left\lbrace 1, 2, \dots , N\right\rbrace $),
		the codeword $X_i=g_i(W,\widetilde{Y}^{i-1},\widehat{h},\bigtriangleup)$, where
		$g_i$ is the transmitter's encoding function, and $\widetilde{Y}^{i-1}=(\widetilde{Y}_1,\dots,\widetilde{Y}_{i-1}) $ is the feedback signal at previous time instants.
		
		\item At time index $i$, the feedback codeword \footnote{Here note that in \cite{JS11},
$h$ is perfectly known by the transceiver, and passive feedback (no feedback encoder) is sufficient for SK-type coding. While in this paper, I-CSIT yields additional error in the SK-type encoding-decoding procedure, hence the active feedback (allowing feedback encoder) is needed to eliminate this type of error.}	$\widetilde{X}_{i}=\widetilde{g}_i(Y^{i},h,\widehat{h},\bigtriangleup,Z^{i-1})$, where $\widetilde{g}_i$ is the receiver's encoding function, ${Y}^{i}=({Y}_1,\dots,{Y}_{i})$ is the output of the forward channel at previous time instants and ${Z}^{i-1}=({Z}_1,\dots,{Z}_{i-1})$ is the QFC noise.
		
		\item At time $N$, the output of the decoder is {\small $\hat{W}=\psi(Y^{N},h,\widehat{h},\bigtriangleup,Z^{N-1})$}, where $\psi$ is the decoding function. The average decoding error probability is defined as
{\small 		\begin{equation}\label{pe}
			P_{e}=\frac{1}{M}\sum\limits_{w = 1}^{M}\Pr\{\hat{W}\neq w|
			w\;\mbox{was sent}\}.
		\end{equation}}
	\end{itemize}
	
	The rate $R$ is said to be $(N,\varepsilon,D)$-achievable if for a given coding blocklength  $N$, error probability $\varepsilon$  and a targeted estimation distortion $D$ about $h$, there exist encoders and decoders such that
	\begin{align}\label{rac}
		\frac{1}{N}\log M\geq {R}-\varepsilon, \,\,\,\, P_{e}\leq \varepsilon, \,\,\,\, \bigtriangleup \leq D.
	\end{align}
	
	The $(N,\varepsilon,D)$-capacity of the model of Figure \ref{fig1} is the supremum over all $(N,\varepsilon,D)$-achievable rates defined above, and it is denoted by $\mathcal{C}_{fd}$.
	
	\subsection{ Main Result}\label{sec5-12}
	\begin{theorem}\label{th1}
		For blocklength $N$, error probability $\varepsilon$ and distortion $D$ about $h$, the $(N,\varepsilon,D)$-capacity $\mathcal{C}_{fd}$ of the model of Figure \ref{fig1} is lower bounded by
{\small 		\begin{align}\label{t111}
			&\mathcal{C}_{fd}\ge \mathcal{R}(N,\varepsilon,D)\nonumber\\
			&=\frac{1}{2N}\log \left( 1+\frac{H^{2}\cdot\text{SNR}\cdot(1+H^2\cdot\text{SNR}\cdot\frac{A}{B})^{N-1}}{L}\right),\nonumber
		\end{align}}
		where
{\small 		\begin{eqnarray*}\label{D2}
			&&A=\left(( \sqrt{3\widetilde{P}}-\sigma_z)[Q^{-1}(\frac{p_m}{2})]^{-1}\right) ^2,\nonumber\\
			&&B=\left( \left| \sqrt{3\widetilde{P}}-\sigma_z\right|[Q^{-1}(\frac{p_m}{2})]^{-1} +\sigma_z \right)^2,\nonumber\\
			&&L=\frac{1}{3}\left[Q^{-1}\left(\frac{\varepsilon}{4}\right)\right]^2,\,\,\,\,
			p_m=\frac{\varepsilon}{2(N-1)},\nonumber\\
			&&H=\max(\mid \hat{h}\mid-{D},0).\nonumber
		\end{eqnarray*}
		}
		\begin{IEEEproof}
			See Section \ref{TH1}.
		\end{IEEEproof}
	\end{theorem}	
		%
			%
	
	\subsection{Numerical  results}	
	\vspace{0.01cm}
	\begin{figure}[H]
		\centering
		\includegraphics[scale=0.25]{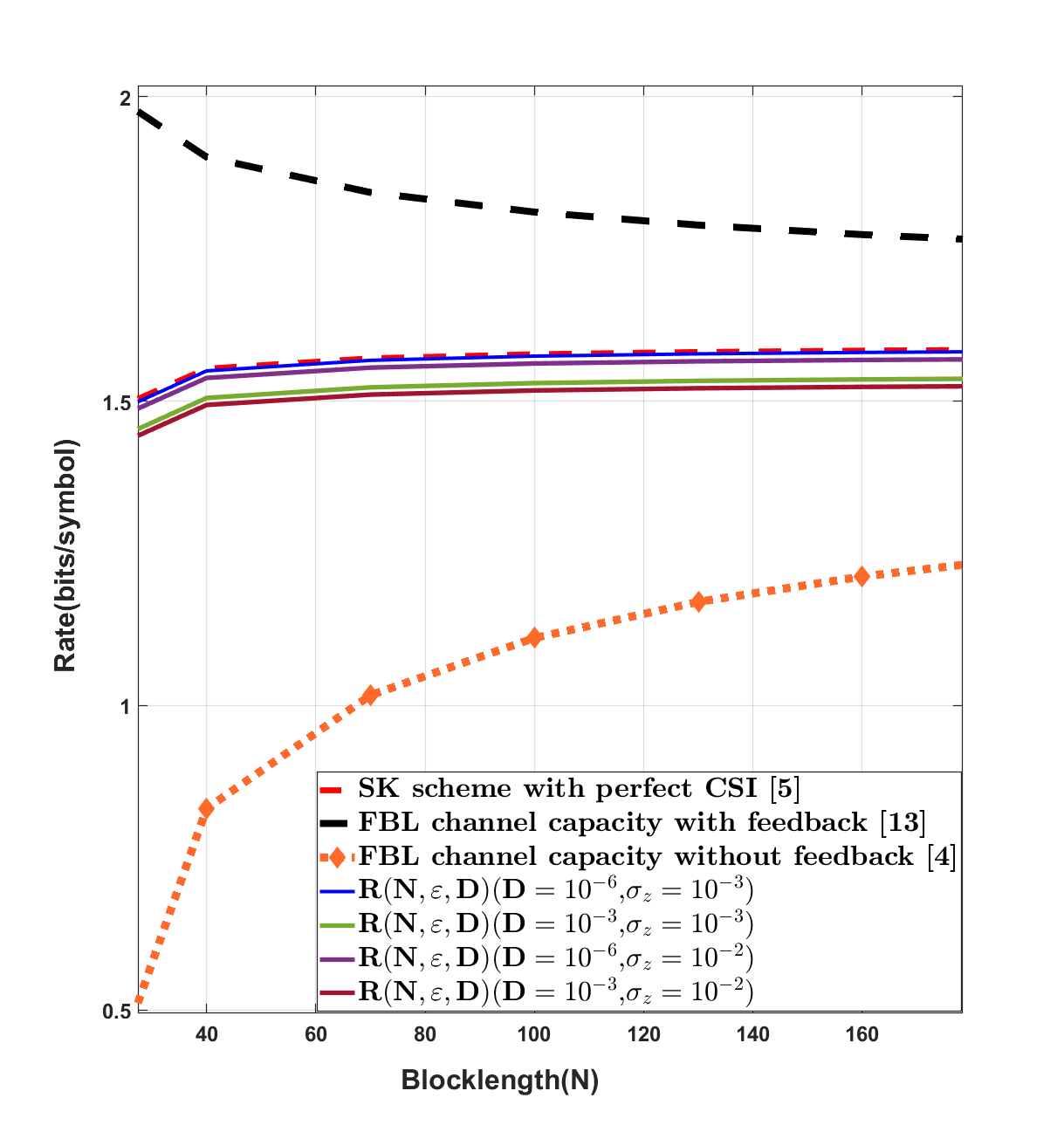}
		\caption{ Achievable rate versus coding blocklength $N$ for $\text{SNR} =10$, $\varepsilon=10^{-6}$, $h=0.9$ and  $\widetilde{P}=10$.}
		\label{fg2}
	\end{figure}	
	\vspace{0.01cm}	
		\begin{figure}[H]
			\centering
			\includegraphics[scale=0.25]{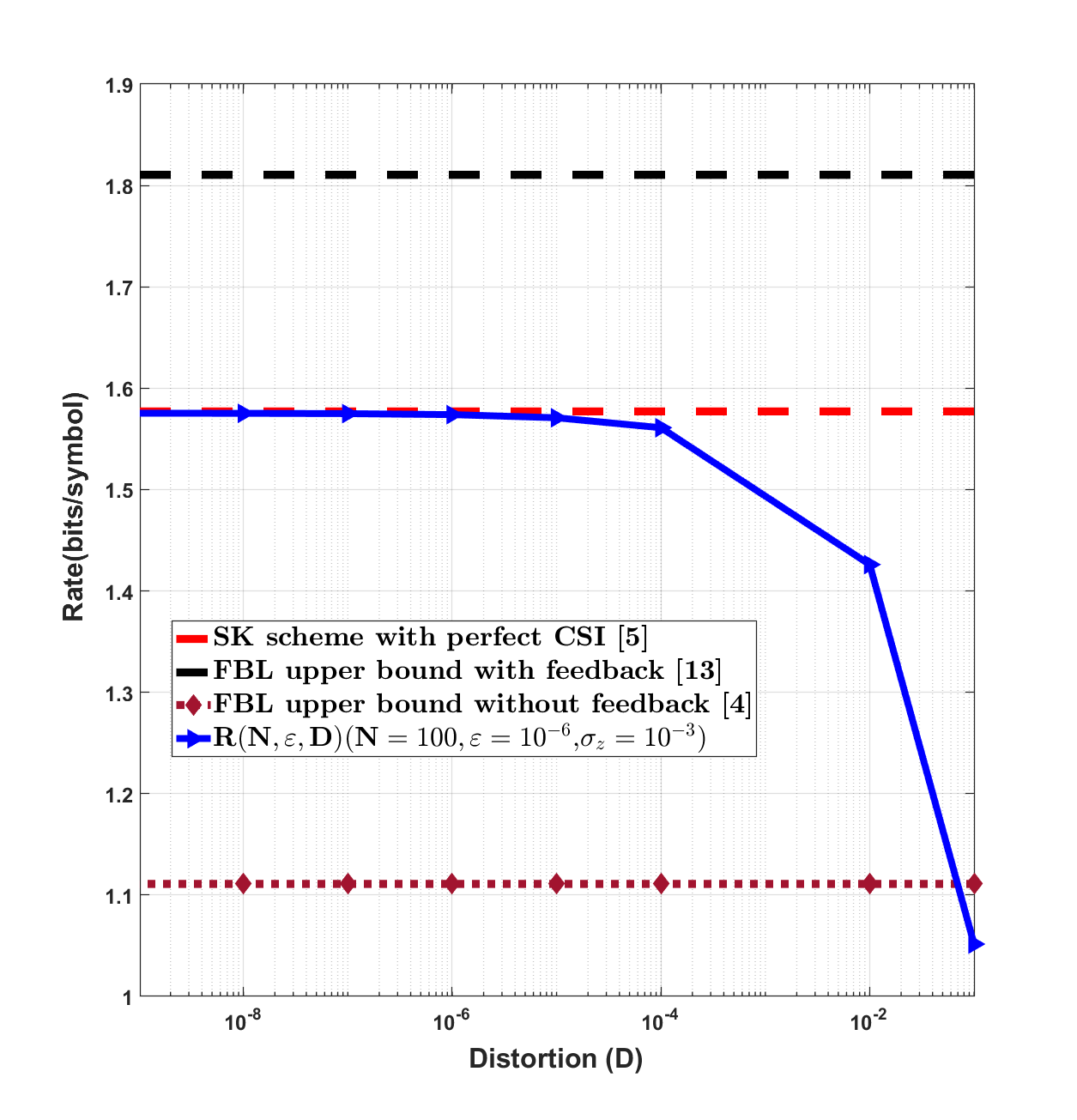}
			\caption{ Achievable rate versus distortion $D$ for $N =100$, $\text{SNR} =10$, $\sigma_z=10^{-3}$, $\varepsilon=10^{-6}$, $h =0.9$ and  $\widetilde{P}=10$.}
			\label{fg3}
		\end{figure}

Both Figures \ref{fg2} and \ref{fg3} show that our proposed scheme almost achieves the rate of the SK scheme for the channel with perfect CSI at both parties when the transmitter's estimation distortion about the CSI and the quantized noise of feedback channel are small. Besides this, both figures show that the quantized feedback combined with I-CSIT still bring FBL rate gain to the quasi-static fading channel with perfect CSI at both parties and without feedback.


		\subsection{ Proof of Theorem \ref{th1}}\label{TH1}
		\renewcommand{\thesubsection}{\Alph{subsection}}

To illustrate the novelty of our scheme compared with existing ones in the literature, the following Figures \ref{in1}-\ref{in3} shows the intuition behind the classical SK scheme, why it is difficult to extend the SK scheme to the channel with I-CSIT and QFC, and how does our scheme work to deal with those problems.

\begin{figure*}[ht!]
	\centering			
	\includegraphics[width=0.68\textwidth]{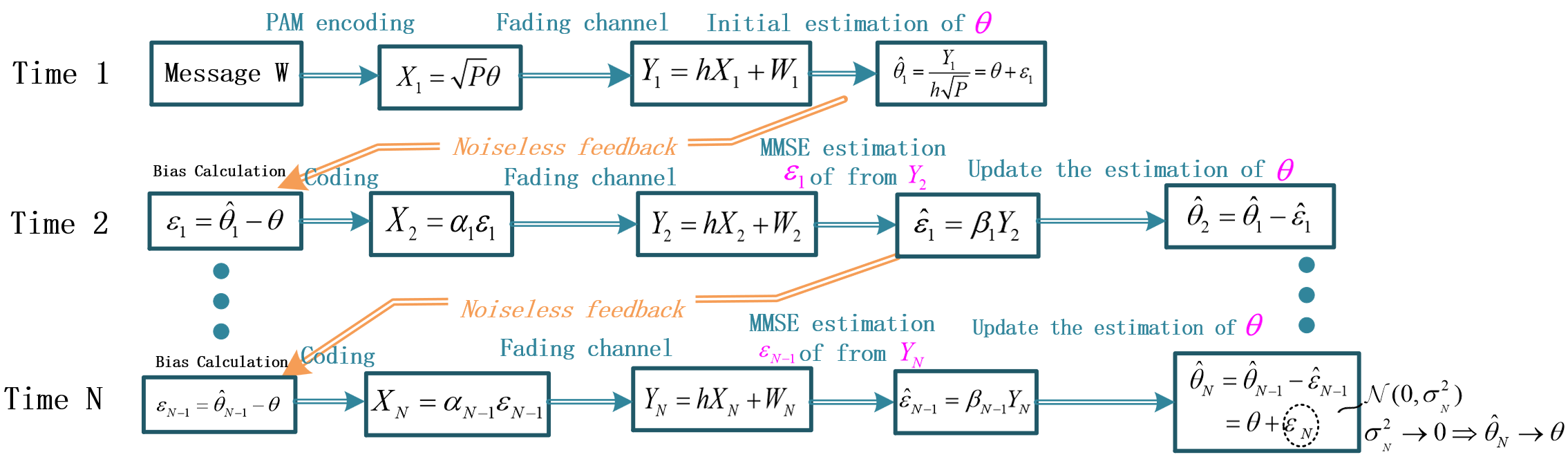}
	\caption{The intuition behind the classical SK scheme}
	\label{in1}
\end{figure*}
	\begin{figure*}
			\centering
			\includegraphics[width=0.68\textwidth]{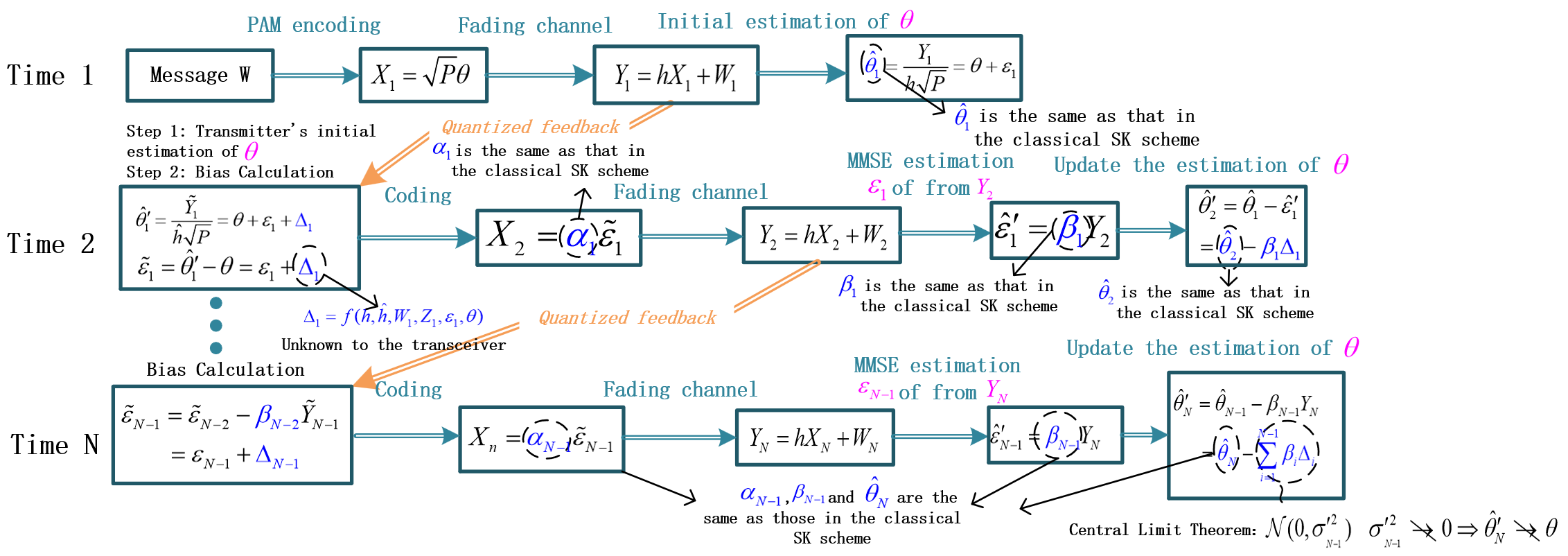}
			\caption{Consequence of directly applying the classical SK scheme to the channel with I-CSIT and quantized feedback}
			\label{in2}
		\end{figure*}
			\begin{figure*}
					\centering
					\includegraphics[width=0.68\textwidth]{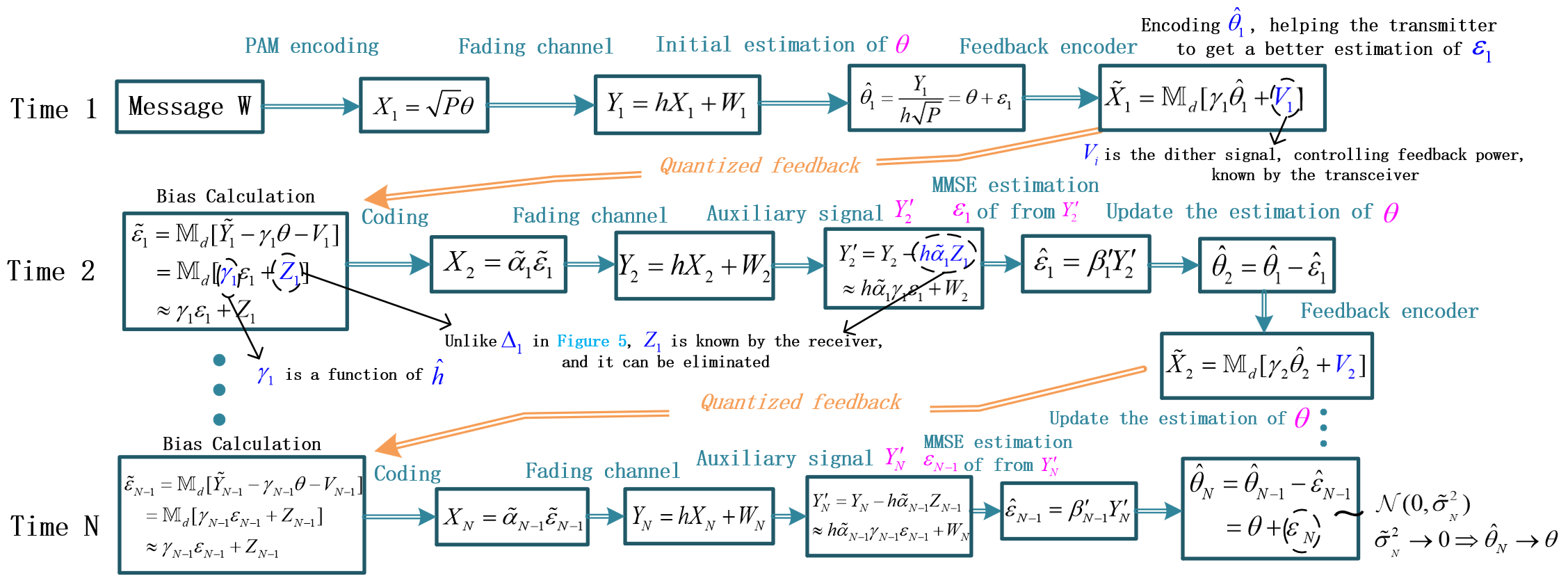}
					\caption{The intuition behind our proposed scheme }
					\label{in3}
				\end{figure*}
As shown in Figure \ref{in2}, directly extending the classical SK scheme to the channel with
I-CSIT and QFC causes an offset which is not known by the receiver, and after several rounds of iteration, the offset is accumulated and cannot be eliminated in the receiver's decoding procedure, which leads to a decoding failure occurs. In Figure \ref{in3}, we show that applying a pair of modulo lattice function based encoder-decoder to the transceiver and introducing an auxiliary signal to the receiver's decoding procedure, the offset caused by I-CSIT and QFC only depends on the quantized noise which is known by the receiver, which indicates that this offset will not be accumulated and can be eliminated by the receiver, and hence the SK scheme still works in such a case.


		
		\subsubsection{Encoding-decoding procedure} \hfil \\		
		$\textbf{Initialization:}$	
		At time instant $1$, first, map the message $M$ to a pulse amplitude modulation (PAM) point $\Theta$, and then
		the transmitter  encodes $\Theta$ as
		\begin{equation}
			X_{1}=\sqrt{{P}}\Theta.
		\end{equation}
		Once receiving
		\begin{equation}
			Y_{1}=hX_{1}+W_{1},
		\end{equation}
		the receiver computes his first estimation $\hat{\Theta}_{1}$ of $\Theta$ by
		\begin{equation}
			\begin{split}
				\hat{\Theta}_{1}&=\frac{Y_{1}}{h\cdot\sqrt{P}}=\Theta+\frac{W_{1}}{h\cdot\sqrt{P}},
			\end{split}
		\end{equation}
		then the receiver sends
		\begin{equation}
			\widetilde{X}_{1}=\mathbb{M}_{\widetilde{d}}[\gamma_{1}\hat{\Theta}_{1}+V_1],
		\end{equation}
		where $\widetilde{d}=\sqrt{12\widetilde{P}}$, $V_{1}$
		is a dither signal uniformly distributed in $[-\frac{\widetilde{d}}{2},\frac{\widetilde{d}}{2})$, $\gamma_{1}$ will be defined later.
		The feedback signal received by the transmitter is denote by
		\begin{equation}
			\widetilde{Y}_{1}=\widetilde{X}_{1}+Z_{{1}}.
		\end{equation}
		At time instant $2$, the transmitter sends
		{	\begin{equation}
				\begin{split}
					X_{2}=\alpha\mathbb{M}_{\widetilde{d}}[\widetilde{Y}_{1}-\gamma_{1}\Theta-V_1]
					\mathop{=}\limits^{\text{(a)}}\alpha\mathbb{M}_{\widetilde{d}}[\gamma_{1}\epsilon_{1}+Z_{1}],\label{X1_1}
				\end{split}
		\end{equation}}%
		where (a) follows from the properties of the modulo-lattice function as shown in~\cite{JS10}, $\epsilon_1=\hat{\Theta}_1-\Theta$, and $\alpha$ will be defined later.
		Once receiving
		\begin{equation}
			Y_{2}=hX_{2}+W_{2},
		\end{equation}
		the receiver calculates an auxiliary signal
		{      \begin{equation}
				\dot{Y}_{2}=Y_{2}-h\alpha\cdot Z_{1},
		\end{equation}}%
		and then the receiver updates his estimation  $\hat{\Theta}_{2}$ of $\Theta$ by using this auxiliary signal, i.e.,
		\begin{equation}
			\hat{\Theta}_{2}=\hat{\Theta}_{1}-\beta_{1}\dot{Y}_{2}\label{A},
		\end{equation}
		where $\beta_{1}$ is the coefficient of MMSE and  will be defined later.	The receiver sends
		\begin{equation}
			\widetilde{X}_{2}=\mathbb{M}_{\widetilde{d}}[\gamma_{2}\hat{\Theta}_{2}+V_2].
		\end{equation}
		where $V_{2}$
		is a dither signal defined the same as $V_{1}$, and $\gamma_{2}$ will be defined later.
		
		$\textbf{Iteration:}$	
		At time instant $i \in \left\lbrace {2},\ldots,N\right\rbrace$, after the transmitter receives the feedback $\widetilde{Y}_{i-1}$, he sends
{\small 		 	\begin{equation}
				\begin{split}	
					X_{i}=\alpha \mathbb{M}_{\widetilde{d}}[\widetilde{Y}_{i-1}-\gamma_{i-1}\Theta-V_{i-1}]
					\mathop{=}\limits^{\text{(b)}}\alpha\mathbb{M}_{\widetilde{d}}[\gamma_{i-1}\epsilon_{i-1}+Z_{i-1}],
				\end{split}\label{X1}
		\end{equation}}
		where (b) follows from the properties of the modulo-lattice function as shown in~\cite{JS10}, and $V_{i-1}$
		is a dither signal uniformly distributed in $[-\frac{\widetilde{d}}{2},\frac{\widetilde{d}}{2})$.
		
		Once receiving
		\begin{equation}
			Y_{i}=hX_{i}+W_{i},
		\end{equation}
		the receiver calculates the auxiliary signal
		{ 		\begin{equation}
				\dot{Y}_{i}=Y_{i}-h\alpha\cdot Z_{i-1}.\label{Y1}
		\end{equation}}%
		Then receiver updates his estimation  $\hat{\Theta}_{i}$ of $\Theta$ by
		\begin{equation}
			\hat{\Theta}_{i}=\hat{\Theta}_{i-1}-\beta_{i-1}\dot{Y}_{i}\label{e1}.
		\end{equation}
		Define $\epsilon_{i}=\hat{\Theta}_{i}-\Theta$, then (\ref{e1}) yields that
		\begin{equation}
			\epsilon_{i}=\epsilon_{i-1}-\beta_{i-1}\dot{Y}_{i}.
		\end{equation}
		
		The receiver sends $\widetilde{X}_i$ back to the transmitter via the QFC, where
		\begin{equation}
			\widetilde{X}_i=\mathbb{M}_{\widetilde{d}}[\gamma_{i}\hat{\Theta}_{i}+V_{i}].
		\end{equation}
		
		$\textbf{Decoding:}$
		At time $N$,  the receiver decodes the message using a minimum distance decoder for $\widehat{\Theta}_{N}$ with respect to the PAM constellation.

		\subsubsection{ Performance analysis} \hfil
		
This subsubsection presents the parameter determination methodology for the proposed scheme and outlines the proof steps for Theorem~\ref{th1}.

The modulo operations of our scheme pose significant challenges to direct decoding error probability analysis. To address this limitation, similar to that of \cite{JS10}, we construct a coupled system that preserves all characteristics of the original system while eliminating the modulo operations. All signals and events in this coupled system are consistently denoted by adding a prime symbol ($'$) to the original notations.
		%
		
		For $i \in \left\lbrace {1},\ldots,N-1\right\rbrace$, define $E_i$ as the event where a modulo-aliasing error occurs, i.e
		{\small 	\begin{eqnarray}\label{Ei_1}
				E_i\mathop{=}\limits^{\rm{def}}\left\lbrace \gamma_{i}\epsilon_{i}+Z_{i}\notin [-\frac{\widetilde{d}}{2},\frac{\widetilde{d}}{2})\right\rbrace.
		\end{eqnarray}}%
		Furthermore, we define $E_{N}$ as the PAM decoding error event
		\begin{eqnarray}
			E_{N}\mathop{=}\limits^{\rm{def}}\left\lbrace \epsilon_{N} \notin [-\frac{{d_{\min}}}{2},\frac{{d_{\min}}}{2})\right\rbrace,
		\end{eqnarray}%
		where $d_{\min}$ is the minimal distance of the PAM.
		
		\begin{lemma}\label{l1}
			For any $N\geq 1$:
{\small 			\begin{equation}\label{model}
				\Pr\left\{\bigcup_{n=1}^{N}E_n \right\}= \Pr\left\{\bigcup_{n=1}^{N}{E_n}^{'} \right\}.
			\end{equation}}
		\end{lemma}
		
		\begin{IEEEproof}
			See proof of Lemma \ref{l1} in Appendix \ref{a-1}.
		\end{IEEEproof}
		Lemma \ref{l1}
		indicates that the error probability in the original system can be bounded by ${E_i}^{'}$ in the coupled system. We obtain
{\small 		\begin{align}\label{E12}
			P_{e}&\leq \Pr\left\{\bigcup_{i=1}^{{N}}{E_i}^{'} \right\}\leq \sum_{i=1}^{{N}} \Pr\left\lbrace E_i^{'}\right\rbrace.
		\end{align}}
		
		It remains to determine the parameters of our scheme. Specifically, recall that
		$\beta_{i}$  is the MMSE coefficient used to estimate $\epsilon_{i}^{'}$ from $\dot{Y}_{i+1}^{'}$. From (\ref{X1}) and (\ref{Y1}), we obtain
		\begin{equation}\label{model}
			\dot{Y}_{i+1}^{'}= h\alpha\gamma_{i}{\epsilon}_i^{'}+W_{i+1},
		\end{equation}
		and solving the optimization for $\beta_{i}$ yields
{\small 		\begin{equation}\label{model}
			\beta_{i}=\frac{E(\epsilon_{i}^{'}\dot{Y}_{i+1}^{'})}{E(\dot{Y}_{i+1}^{'2})}.
		\end{equation}}
		
		Observing that $\epsilon_{n+1}^{'}=\epsilon_{n}^{'}-\widehat{\epsilon_{n}^{'}}$, and using $\beta_{i}$ defined above, we obtain a recursive formula
		for $\sigma_i^{2}$, which is given by
		{ 	\begin{align}
				\sigma_{i+1}^{2}=
				(1+\frac{h^{2}\alpha^2\gamma_i^2\sigma_i^{2}}{\sigma^2})^{-1}\sigma_i^{2},\,\,\,\,
				\sigma_1^{2}=\frac{\sigma^2}{Ph^2},
		\end{align}}
		where $\sigma_i^{2}=E({\epsilon}_i^{'})^2$.

Here note that $\left\lbrace \sigma_i^{2}\right\rbrace $ depends on $h$, which is not known by the transmitter. Hence the transmitter can only adopt $\left\lbrace \sigma_i^{2}(H)\right\rbrace $ instead of $\left\lbrace \sigma_i^{2}\right\rbrace $ to the encoding procedure, where
{\small \begin{align}
			\sigma_{i+1}^{2}(H)=(1+\frac{H^{2}\alpha^2\gamma_i^2\sigma_i^{2}(H)}{\sigma^2})^{-1}\sigma_i^{2}(H),\,\,
		\sigma_1^{2}(H)=\frac{\sigma^2}{PH^2},\nonumber
		\end{align}}%
		and $H=\max(\mid \hat{h}\mid-{D},0)$. By the triangle inequality, we can easily check that $\mid h\mid\geq H$, which leads to the actual transmitting power is smaller than the power constraint.
				\begin{lemma}\label{l2}
			For any $i>1$, $\sigma_i^{2}\leq\sigma_i^{2}(H)$.
		\end{lemma}
		\begin{IEEEproof}
			See proof of Lemma \ref{l2} in Appendix \ref{a-2} .
		\end{IEEEproof}

		
		Letting $\Pr\left\lbrace E_{N}^{'}\right\rbrace=\frac{\varepsilon}{2} $, $\Pr\left\lbrace E_1^{'}\right\rbrace = \cdots = \Pr\left\lbrace E_{N-1}^{'}\right\rbrace=\frac{\varepsilon}{2(N-1)} =p_m$, and using the definition of the event $E_i^{'}$ in (\ref{Ei_1}) and $\widetilde{d}=\sqrt{12\widetilde{P}}$, we conclude that
{\small 		\begin{equation}
			\gamma_i=\sqrt{\frac{A}{\sigma^{2}_i(H)}},
		\end{equation}}%
		 where {\small $A\mathop{=}\limits^{\text{def}}\left(( \sqrt{3\widetilde{P}}-\sigma_z)[Q^{-1}(\frac{p_m}{2})]^{-1}\right) ^2$}.
%
          The detailed derivation of $\gamma_i$ refers to Appendix \ref{a-3}.

The parameter $\alpha$ in (\ref{X1}) ensures the transmit signal power does not exceed the power constraint $P$. Specifically, $\alpha$ is given by
{\small 	   \begin{align}
					\alpha=\sqrt{\frac{P}{B}},
		\end{align}}%
		where {\small $	B\mathop{=}\limits^{\text{def}}\left( \left| \sqrt{3\widetilde{P}}-\sigma_z\right|[Q^{-1}(\frac{p_m}{2})]^{-1} +\sigma_z \right)^2$}.
%
	The detailed derivation of $\alpha$ refers to Appendix \ref{a-4}.
	
		By substituting $\alpha$ and $\gamma_i$  back into $\beta_i$ and $\sigma_i^{2}(H)$,  we obtain
 		\begin{align}
			&\beta_{i}=\frac{h\sqrt{\frac{P}{\sigma_{i}(H)^{2}}\cdot\frac{A}{B}}\cdot\sigma_{i}^{2}}{h^{2}{\frac{P}{\sigma_{i}(H)^{2}}\cdot\frac{A}{B}}\cdot\sigma_{i}^{2}+\sigma^{2}},\\
			&\sigma_i^{2}(H)	=\frac{1}{{H}^2\text{SNR}}\left( \frac{1}{1+{H}^2\text{SNR}\frac{A}{B}}\right) ^{i-1}.
		\end{align}
		

				\renewcommand{\thesubsection}{\Alph{subsection}} 		
						
						Finally, through the analysis of the decoding error probability in Appendix \ref{a-5}, we conclude that for a given coding blocklength $N$ and error probability $\varepsilon$, the average decoding error probability $P_e$ of the proposed scheme does not exceed $\varepsilon$ by appropriately choosing the parameters $\gamma_{i}$ and $\beta_{i}$. The corresponding achievable rate is given by
						
{\small 						\begin{align}
	 & \mathcal{R}(N,\varepsilon,D)=
							\frac{1}{2N}\log \left( 1+\frac{H^{2}\cdot\text{SNR}(1+H^2\cdot\text{SNR}\cdot\frac{A}{B})^{N-1}}{L}\right),\nonumber
\end{align}}
						where $L=\frac{1}{3}\left[Q^{-1}\left(\frac{\varepsilon}{4}\right)\right]^2$, which completes the proof.
						

\section{Conclusion}\label{sec5}
This paper proposes an efficient SK-type coding scheme for the quasi-static fading channel with I-CSIT and QFC, and establishes the rate-CSI estimation distortion tradeoff under given coding blocklength and decoding error probability.
Numerical results show that the quantized feedback combined with I-CSIT still bring FBL rate gain to the quasi-static fading channel with perfect CSI at both parties and without feedback.
													
																	

%

\renewcommand{\theequation}{A\arabic{equation}}
\section*{Appendix}\label{P}
\setcounter{equation}{0}
\subsection{Proof of Lemma \ref{l1}}\label{a-1}

%
%
Define the event
\begin{equation}\label{model}
	\begin{array}{l}
		J_n\mathop{=}\limits^{\rm{def}} \bigcap\limits_{i=1}^{n}\overline{E_i},\nonumber
	\end{array}
\end{equation}
where $\overline{E_i}$ represents the complement of $E_i$.

Let us show by induction that $J_N=J^{'}_N$.  For $n=1$, we have
{ 		{ 		\begin{equation}\label{dasb1}	
			\begin{split}	
				&J_{1}=\overline{E_{1}} \\
				&=\left\lbrace \gamma_{1}\epsilon_1+Z_{1}\in [-\frac{\widetilde{d}}{2},\frac{\widetilde{d}}{2})\right\rbrace \\
				&=\left\lbrace \gamma_{1}\epsilon^{'}_{1}+Z_{1}\in [-\frac{\widetilde{d}}{2},\frac{\widetilde{d}}{2})\right\rbrace\\
				&=J_{1}^{'},
			\end{split}			
\end{equation}}}
where (\ref{dasb1}) follows from the sample path identity. 

Assuming	$J_{N-1}=J^{'}_{N-1}$ and using the sample path identity again, we have
{	\begin{equation}
	\begin{split}		
		&J_{N}
		=\lbrace \gamma_{N}\epsilon_{N-1}+Z_{N}\in [-\frac{\widetilde{d}}{2},\frac{\widetilde{d}}{2})\rbrace \bigcap J_{N-1}\\
		&	=\lbrace \gamma_{N}\epsilon^{'}_{N}+Z_{N}\in [-\frac{\widetilde{d}}{2},\frac{\widetilde{d}}{2})\rbrace \bigcap J_{N-1}^{'}\\
		&=J_{N}^{'}.
	\end{split}
	\end{equation}}
	Analogously, we conclude that $J_{n-1} \bigcap E_n=J_{n-1}^{'} \bigcap E_n^{'}$. Then we have
	\begin{equation}\label{model}
\begin{array}{l}
	\Pr\left\{\bigcup\limits_{n=1}^{N}E_n  \right\}\\
	=\Pr\left\lbrace E_{1}\right\rbrace +\sum\limits_{n=2}^{N} \Pr\left \{ \bigcap\limits_{i=1}^{n-1}\overline{E_i}\bigcap E_n\right \}\\
	=\Pr\left\lbrace \overline{J_{1}}\right\rbrace +\sum\limits_{n=2}^{N} \Pr\left \{ J_{n-1}\bigcap E_n\right \}\\
	=\Pr\left\lbrace \overline{J^{'}_{1}}\right\rbrace +\sum\limits_{n=2}^{N} \Pr\left \{ J^{'}_{n-1}\bigcap E^{'}_n\right \}\\
	=\Pr\left\{\bigcup\limits_{n=1}^{N}{E_n}^{'}  \right\}.
\end{array}
\end{equation}

\subsection{Proof of Lemma \ref{l2}}\label{a-2}																														

For $i=1$, we have
\begin{equation}
\begin{array}{l}
	\sigma_1^{2} =\frac{\sigma^2}{Ph^2}\leq \frac{\sigma^2}{PH^2}=\sigma_1^{2}(H).\nonumber
\end{array}
\end{equation}
Then assuming $\sigma_{i-1}^{2}\leq\sigma_{i-1}^{2}(H)$, we have
{ 														\begin{align}\label{AA}
	\sigma_i^{2} &=(1+\frac{h^{2}\alpha^2\gamma_{i-1}^2\sigma_{i-1}^{2}}{\sigma^2})^{-1}\sigma_{i-1}^{2}\nonumber\\
	&=({\frac{1}{\sigma_{i-1}^{2}}+\frac{h^{2}\alpha^2\gamma_{i-1}^2}{\sigma^2}})^{-1}\nonumber\\
	&\leq (\frac{1}{\sigma_{i-1}^{2}(H)	}+\frac{H^{2}\alpha^2\gamma_{i-1}^2}{\sigma^2})^{-1}\nonumber\\
	&=\sigma_i^{2}(H).\nonumber
	\end{align}}
	
	\subsection{Determination of $\gamma_i$}\label{a-3}	
	Letting { $\Pr\left\lbrace E_{N}^{'}\right\rbrace=\frac{\varepsilon}{2} $, $\Pr\left\lbrace E_1^{'}\right\rbrace = \cdots = \Pr\left\lbrace E_{N-1}^{'}\right\rbrace=\frac{\varepsilon}{2(N-1)} =p_m$}, and using the definition of the event $E_i^{'}$ in (\ref{Ei_1}) and $\widetilde{d}=\sqrt{12\widetilde{P}}$, we conclude that
	\begin{align}				
\Pr\left\lbrace E_i^{'}\right\rbrace  &=\Pr\left\{\gamma_i{\epsilon}_i^{'}+Z_i\notin\left[-\frac{\widetilde{d}}{2},\frac{\widetilde{d}}{2}\right)\right\} \nonumber\\
&=\Pr\left\{\arrowvert\gamma_i{\epsilon}_i^{'}+Z_i\arrowvert\geq\sqrt{3\widetilde{P}}\right\}\nonumber\\
&\mathop{\leq}\limits^{(a)} \Pr\left\{\arrowvert\gamma_i{\epsilon}_i^{'}\arrowvert+\sigma_z\geq\sqrt{3\widetilde{P}}\right\}\nonumber\\
&= 2Q\left(\frac{\sqrt{3\widetilde{P}}-\sigma_z }{ \sqrt{E\left( \gamma_i{\epsilon}_i^{'}\right)^{2}}} \right) \nonumber\\
&\mathop{\leq}\limits^{(b)} 2Q\left(\frac{\sqrt{3\widetilde{P}}-\sigma_z }{ \sqrt{\gamma_i^2{\sigma}_i^{2}(H)}} \right)\mathop{=}\limits^{(c)}p_m,									
\end{align}
where (a) follows from the fact that the quantized noise $Z_i$ is upper bounded by $\sigma_z$, (b) follows from  $Q(x)$-function  is monotonically decreasing while $x$ is increasing,
and (c) follows from choosing
\begin{align}
\gamma_i=\sqrt{\frac{A}{\sigma^{2}_i(H)}},
\end{align}
where
\begin{align}
A\mathop{=}\limits^{\text{def}}\left(( \sqrt{3\widetilde{P}}-\sigma_z)[Q^{-1}(\frac{p_m}{2})]^{-1}\right) ^2.
\end{align}

\subsection{Determination of $\alpha$}\label{a-4}	

We choose $\alpha$ to ensure the actual transmitting power does not exceed the average power constraint $P$, namely,	
\begin{eqnarray}\label{dsb-1}
&&E(X_{i+1})^2=E(\alpha(\gamma_i\epsilon^{'}_i+Z_i))^2\nonumber\\
&&=\alpha^2E(\gamma_i\epsilon^{'}_i+Z_i)^2=\alpha^2\|\gamma_i\epsilon^{'}_i+Z_i\|_R^2\nonumber\\
&&\mathop{\leq}\limits^{(a)}\alpha^2(\|\gamma_i\epsilon^{'}_i\|_R+\|Z_i\|_R)^2\nonumber\\
&&=\alpha^2(\sqrt{\gamma_i^2 \sigma_i^2}+\sigma_z)^2\nonumber\\
&&\leq \alpha^2(\sqrt{\gamma_i^2 {\sigma}_i^{2}(H)}+\sigma_z)^2\mathop{=}\limits^{(b)}P,
\end{eqnarray}
where $\|X\|_R\mathop{=}\limits^{\rm{def}}(E(X^2))^{1/2}$, (a) follows from the Minkowski inequality, and (b) follows from choosing
{        { 	\begin{align}
		\alpha=\sqrt{\frac{P}{B}},
		\end{align}} }
		where
		\begin{align}
B\mathop{=}\limits^{\text{def}}\left( | \sqrt{3\widetilde{P}}-\sigma_z|[Q^{-1}(\frac{p_m}{2})]^{-1} +\sigma_z \right)^2.
\end{align}

\subsection{Decoding error probability analysis}\label{a-5}	
According to the parameters given above, we have
\begin{equation}\label{model}
\text{SNR}_i=\frac{1}{\sigma_{i}^{2}}, \ \	\text{SNR}_i(H)=\frac{1}{\sigma_{i}^{2}(H)},
\end{equation}
where $\text{SNR}_i\geq \text{SNR}_i(H)$ for any $i\geq 1$. Then (\ref{E12}) can be re-written by
{ 	\begin{align}
	P_{e}	&\leq  \sum_{i=1}^{N} \Pr\left\lbrace E_i^{'}\right\rbrace \nonumber\\
	&\leq (N-1)p_m+\Pr\left\lbrace E_{N}^{'}\right\rbrace \nonumber\\
	&\leq \frac{\varepsilon}{2}+\Pr\left\lbrace E_{N}^{'}\right\rbrace \nonumber\\
	&\mathop{\leq}\limits^{(a)}\frac{\varepsilon}{2}+2Q\left( \sqrt{\frac{3\cdot\text{SNR}_{N}}{2^{2N{R}}-1}}\right) \nonumber\\
	&\leq\frac{\varepsilon}{2}+2Q\left( \sqrt{\frac{3\cdot\text{SNR}_{N}}{2^{2N{R}}}}\right)\nonumber\\
	&\leq\frac{\varepsilon}{2}+2Q\left( \sqrt{\frac{3\cdot\text{SNR}_{N}(H)}{2^{2N{R}}}}\right)\mathop{=}\limits^{(b)}\varepsilon,
	\end{align}}%
	where (a) follows from the detection error probability of PAM in \cite{JS10}, and (b) follows from choosing
	\begin{equation}\label{model}
Q\left( \sqrt{\frac{3\cdot\text{SNR}_{N}(H)}{2^{2N{R}}}}\right)=\frac{\varepsilon}{4},
\end{equation}
which indicates that
{ 	\begin{align}
	& \mathcal{R}(N,\varepsilon,D)=\frac{1}{2N}\log\left(\frac{\text{SNR}_{N}(H)}{L} \right) \nonumber \\
	& =\frac{1}{2N}\log \left( \frac{H^{2}\text{SNR}(1+H^2\text{SNR}\frac{A}{B})^{N-1}}{L}\right),
	\end{align}}
	where
	$L=\frac{1}{3}\left[Q^{-1}\left(\frac{\varepsilon}{4}\right)\right]^2$.

														\end{document}